# Growth, Electronic Structure and Superconductivity of Ultrathin Epitaxial CoSi$_2$ Films

Yuan Fang[1,2,#], Ding Wang[1,2,#], Peng Li[1,2], Hang Su[1,2], Tian Le[1,2], Yi Wu[1,2], Guo-Wei Yang[1,2], Hua-Li Zhang[1,2], Zhi-Guang Xiao[1,2], Yan-Qiu Sun[2], Si-Yuan Hong[2], Yan-Wu Xie[2], Huan-Hua Wang[3], Chao Cao[4], Xin Lu[1,2,5], Hui-Qiu Yuan[1,2,5], Yang Liu[1,2,5*]

[1]Center for Correlated Matter, Zhejiang University, Hangzhou, China

[2]Zhejiang Province Key Laboratory of Quantum Technology and Device, Department of Physics, Zhejiang University, Hangzhou, P. R. China

[3]Institute of High Energy Physics, Chinese Academy of Sciences, Beijing 100049, China

[4]Department of Physics, Hangzhou Normal University, Hangzhou, China

[5]Collaborative Innovation Center of Advanced Microstructures, Nanjing University, Nanjing, China

\# these authors contribute equally to this work

\* Email: yangliuphys@zju.edu.cn

**Abstract**

We report growth, electronic structure and superconductivity of ultrathin epitaxial CoSi$_2$ films on Si(111). At low coverages, preferred islands with 2, 5 and 6 monolayers height develop, which agrees well with the surface energy calculation. We observe clear quantum well states as a result of electronic confinement and their dispersion agrees well with density functional theory calculations, indicating weak correlation effect despite strong contributions from Co 3$d$ electrons. *Ex-situ* transport measurements show that superconductivity persists down to at least 10 monolayers, with reduced $T_c$ but largely enhanced upper critical field. Our study opens up the opportunity to study the interplay between quantum confinement, interfacial symmetry breaking and superconductivity in an epitaxial silicide film, which is technologically relevant in microelectronics.

1. Introduction

In a thin film with the thickness smaller than the electron coherence length (typically a few tens or hundreds of nm), electron waves bounce back and forth between the surface and the interface, forming the so-called quantum well states (QWS) [1][2][3][4]. Accordingly, the electronic states are quantized with discretely allowed $k_z$'s (wave vector along $z$). As the film thickness grows, new subbands emerge periodically below the Fermi level ($E_F$), which in turn gives rise to thickness-dependent density of states (DOS) at $E_F$. Since many physical properties of the films, including thermal stability, superconductivity, magnetism, etc., depend critically on the DOS at $E_F$ [5][6], thickness-dependent oscillation of these properties can be expected. Based on the standard Bohr-Sommerfeld quantization model [2], the evolution of quantum well states with film thickness will produce oscillations in DOS at $E_F$ with a period around half of the Fermi wavelength $\lambda_F = 2\pi/k_F$, where $k_F$ is the Fermi vector of the band structure along $z$ [7].

While many previous studies of quantum size effect (QSE) are focused on elemental metal systems, here we report a silicide film grown epitaxially on Si(111), $CoSi_2$. With its good thermal stability and low resistivity, $CoSi_2$ is actually a commonly used silicide in CMOS technology [8]. Another important advantage is that the $CoSi_2$ lattice is well matched to Si (mismatch < 1.2%), facilitating scalable growth of epitaxial films. Bulk $CoSi_2$ is also a superconductor with $T_c \approx 1.2$ K and the upper critical field is 105 Oe [9][10]. Epitaxial $CoSi_2$ films were also reported to be superconducting above a critical thickness of about 10 nm [11][12]. Recent research reported ultralow $1/f$ noise in superconducting epitaxial $CoSi_2$ films grown on Si, implying possible applications for developing quiet quantum qubits and scalable superconducting circuits for future quantum computing [13]. In addition, the electronic states of CoSi$_2$ near $E_F$ contain substantial contribution from Co 3$d$ electrons and the study of the QWSs in the ultrathin limit would allow us to explore the interplay between possible electron correlation and quantum confinement [14][15][16]. These interesting properties motivate us to systematically study the growth and electronic properties of CoSi$_2$ films.

The CoSi$_2$ films grown on Si(111) have been studied before. Studies from scanning tunneling microscope (STM) have found evidence of preferred (also called "magic") thickness for CoSi$_2$ films on Si (111) [17]. DFT calculations predict a trilayer oscillation of surface energy as a function of film thickness [18]. However, the evolution of the momentum-resolved electronic structure with thickness remains elusive experimentally. Here we report direct observation of QWSs in ultrathin CoSi$_2$ films on Si (111) by angle resolved photoemission spectroscopy (ARPES), providing direct support for the preferred magic island growth predicted by density functional calculations (DFT). Our APRES results agree very well with the calculated band structure from DFT, indicating very small correlation effect in this compound despite strong spectral contributions from Co 3$d$ electrons. We also show that the superconductivity persists down to 10 monolayers (MLs), with reduced $T_c$, but the upper critical field is much larger compared to bulk. Our results open up the opportunity to study the interplay between quantum confinement and superconductivity in this technologically important silicide.

2. **Experimental and Computational Details**

All CoSi$_2$ films were grown by molecular beam epitaxy (MBE) in a growth chamber with a base pressure of 1.8 x 10$^{-10}$ mbar. High purity Co was deposited onto the Si(111) substrate to react with the Si and form epitaxial CoSi$_2$ films. The deposition of Co was done with a high-temperature effusion cell (set at 1200°C) to yield a rate of approximately 0.37 ML per minute, as determined by a quartz crystal monitor (QCM) mounted near the sample. The Si substrates were first heated to 1000°C to form 7x7 reconstruction, and the substrates were maintained at 300°C during the Co deposition. After deposition, the films were annealed to 500°C-600°C to form well-ordered CoSi$_2$ films. The film quality is monitored in real time by high energy electron diffraction (RHEED). ARPES measurements were performed immediately after the film growth, by transferring the sample under ultrahigh vacuum from the MBE growth chamber to the connected ARPES chamber. A five-axis manipulator cooled by a closed-cycle helium refrigerator was employed for band structure measurements and sample cooling. All APRES spectra were taken with a sample temperature of ~15 K, above the superconducting transition temperature (Tc < 1.1 K). All the ARPES data were taken with He-I photons (21.2 eV). The overall energy (momentum) resolution is ~10 meV (~0.01 Å$^{-1}$). Synchrotron X-ray scattering experiments (Fig. 1(c,d)) were performed at 1W1A Diffuse X-ray Scattering Station, Beijing Synchrotron Radiation Facility (BSRF-1W1A). The wavelength of incoming X-ray is set to 1.545 Å.

Electronic structure calculations were performed using DFT based on the projected augmented wave method, as implemented in the Vienna ab initio simulation package (VASP) [19]. Perdew, Burke and Enzerhoff flavor of exchange-correlation functional was employed [20]. An energy cutoff of 400 eV was employed to converge the bulk calculation to 1 meV/atom. The k mesh is 15 × 15 × 1 for slab calculation and 16 × 16 × 16 for projected bulk band calculation. For the slab calculations with additional Si bilayer termination and bulk-like termination, the in-plane lattice constants are taken to be the same as Si and structure relaxations are performed by relaxing the outmost 5 atomic layers and 3 atomic layers, respectively.

The electrical resistance measurements were carried out using a Quantum Design Physical Property Measurement System (PPMS-9T) equipped with He$^3$ option. The four-contact method was used for the resistance measurements. The applied magnetic field is along the $z$ direction.

## 3. Results and Discussion

### 3.1. Film growth and oscillatory surface energy

Fig 1.(a) shows the RHEED patterns before and after depositing 5ML CoSi$_2$ film. Here the film thickness refers to the estimated nominal coverage of the CoSi$_2$ film, based on the deposited amount of Co measured from QCM (error bar is typically <10%), i.e., 1 ML = 0.86 Å of Co. Before growth, the RHEED shows clear Si 7×7 reconstruction; after film deposition and annealing, sharp 1×1 streaky patterns develop corresponding to epitaxial CoSi$_2$ films. Our optimal annealing condition is determined by comparing the ARPES spectra at different annealing stages. Fig1(b) shows the atomic force microscopy (AFM) image of a 20ML CoSi$_2$ film. While the AFM image shows that a majority of the sample surface is covered by uniform CoSi$_2$ islands with large areas (light color), there also exist deep holes (dark color) between these flat CoSi$_2$ islands. Fig. 1(c) and 1(d) are synchrotron X-ray scattering measurement results of a 100 ML CoSi$_2$ film. Here the H, K and L directions are defined to be along the bulk [$2\bar{1}\bar{1}$], [$11\bar{2}$] and [111] directions in the substrate (Si) reciprocal space, respectively. From the specular (H,K,L)=(0,0,L) scan shown in Fig. 1(c), the lattice constant of CoSi$_2$ film along the z direction can be determined to be 3.088 Å, which is slightly smaller than the bulk value (3.092 Å) due to strain effect. The epitaxial relation between the film and substrate can be directly verified by the off-specular scans shown in Fig. 1(d), where the Bragg peaks from the Si substrate and the CoSi$_2$ film can be simultaneously observed. The small peak located at L=3.5 in (2,0,L) scan is likely from scattering of second harmonic photons. It is interesting to note that for (0,2,L) and (2,-2,L) scans (which are identical due to three-fold symmetry of the system), the observed substrate (film) peaks are at L~4 (L~5). However, for the (2,0,L) scan (which is 60° rotated from the other two scans), the reverse is observed: the substrate (film) peaks are at L~5 (L~4). This indicates that the lattice of CoSi$_2$ film is rotated 60° along z direction with respect to that of Si substrate, which is consistent with a previous Auger study [21].

Fig 2.(a,c) shows the calculated surface energy of freestanding CoSi$_2$ films from DFT (red curves), for bulk-like termination (a) and those terminated by additional Si bilayers on both sides (c). Here we include the calculations for double Si layer termination because such termination was implied by a few previous studies [21][22][23][24] and indeed supported by our detailed comparison between experiments and calculations, which we shall discuss more below. To obtain the surface energy, we first calculate the total energy of freestanding films with various thicknesses. Then we fit the total energies of three thickest films (8, 9, 10 ML) using a linear function. Finally, by subtracting this linear function, we could obtain the thickness-dependent surface energy [18][25]. The surface energy can be conventionally fitted by the following functional form [5][26]:

$$E_s(N) = B \frac{\sin(2k_F Nt + \varphi)}{N^\alpha} + C, \tag{1}$$

where $k_F$ is the Fermi wave vector along the Γ-L direction, $t$ is the monolayer thickness, $B$ is an amplitude parameter, $\varphi$ is the phase shift that depends on the interface properties, $\alpha$ is a decay exponent, and $C$ is a constant offset. The fits are shown as blue dashed lines in (a,c). We obtain $k_F = 0.335$ Å$^{-1}$ and $k_F = 0.349$ Å$^{-1}$ for the bulk-like termination and Si bilayer termination, respectively. These fitted $k_F$'s are close to $k_F = 0.368$ Å$^{-1}$ from DFT calculations [18], which implies a trilayer oscillation of the surface energy governed by the bulk band structure. Indeed, both terminations show damped oscillation in a trilayer periodicity, although the oscillation phases ($\varphi'$s) are apparently different. The trilayer oscillation of the surface energy can be better illustrated by the second derivative of surface energy shown in Fig 2.(b,d) [27]:

$$\Delta^2 E_s(N) = E_s(N+1) + E_s(N-1) - 2E_s(N). \qquad (2)$$

Here the positive values of second derivative of surface energy indicate stable thicknesses. Calculation for the bulk-like termination predicts 1ML, 2ML, 4ML, 5ML, 7ML as the preferred thicknesses, while for Si bilayer termination it predicts 2ML, 5ML, 6ML, 8ML as the preferred thicknesses. As we shall show below, our experimental observations agree with the prediction from the Si bilayer termination. We also do the surface energy calculation with bulk-like termination using the in-plane lattice constants of bulk CoSi2 (see supplementary information [28] Fig. S2). It gives the same stable thicknesses as previous calculation using in-plane lattice constants of Si substrate. Thus we conclude that the calculated stable thicknesses are robust whether the film shares the same in-plane lattice constant with substrate or adopts its bulk value.

3.2. Electronic Structure of ultrathin CoSi$_2$ films and comparison with DFT

The electronic structure for CoSi$_2$ films with various nominal coverages is summarized in Fig 3(a) (second derivatives of the spectra are shown in supplementary information [28] Fig.

S1). We emphasize again that the coverage here refers to the nominal $CoSi_2$ coverages based on the estimated deposition of Co. At low coverages, one can clearly see a set of hole bands centered at $\bar{\Gamma}$, whose energy separation is dependent on the film coverage. This is the characteristic QWSs due to quantum confinement effect in ultrathin films. Close inspection suggests that 1 and 1.5 ML films actually share similar spectral feature. The spectral shape changes suddenly for 2 ML film. However, for thicker films from 2.5 ML up to 5 ML, the spectra again show similar features. Such spectral evolution with thickness can also be confirmed by the energy distribution curves (EDCs) shown in Fig. 3(b). Since the energy positions of the QWSs are dependent on the film thickness, the above coverage evolution could only be explained by formation of islands with preferred height. In other words, the growth is not ideal layer-by-layer, but develops islands that are energetically more favorable [29][30]. Indeed, as we shall show below by detailed comparison with DFT calculations, the 2 ML island forms preferably below 1.5 ML film coverage, while a mixture of 5 ML and 6 ML islands develops for film coverages between 2.5 ML and 5 ML (see Fig. 3(c)). Such preferred thickness is in excellent agreement with the calculated surface energy shown in Fig. 2(c, d). For films with coverages larger than 8 ML, the discrete QWSs become washed out, probably due to increased film (island) roughness. This is expected since the oscillation of the surface energy quickly damps with increasing thickness, and therefore the thickness preference also diminishes.

Comparisons between ARPES data and DFT calculation are summarized in Fig 4. The films with 1 ML and 3 ML coverages (Fig. 4(a,e)) are chosen here because they consist of the magic islands with 2 ML and 5,6 ML height, respectively. To make the spectra more visible, here we take the second derivatives of the original ARPES data. Fig 4(b,f) show the slab calculation with additional Si bilayers on both sides, for 2 ML (b) and 5,6 ML islands (f). The outer (inner) Si atom in the Si bilayer have the same in-plane coordinate as the Si (Co) atom in the bulk Si-Co-Si layer, consistent with a previous report [21]. The detailed positions are further optimized in the DFT calculation. There is good agreement between experiments (a,e) and calculations (b,f). By contrast, the DFT calculations using the nominal coverage (1 ML and 3 ML) yield very poor agreement with experiments. In addition, Fig 4(c,g) show the slab calculations with bulk-like termination, which apparently deviate considerably with the experimental data, particularly for the 1ML film. Another common approach to quantitatively understand QWSs is through the projected bulk band calculation using the Bohr-Sommerfeld quantization rule [2]:

$$2k_\perp(E)Nt + \varphi_s + \varphi_i = 2n\pi \qquad (3)$$

where $k_\perp(E)$ is the perpendicular momentum as a function of energy $E$ in accordance with the bulk band structure, $N$ is the film thickness in units of ML, $t$ is the thickness of one ML, $\varphi_s$ ($\varphi_i$) is the phase shift at the surface (interface), and $n$ is the quantum number of each QWS subband. The total phase shift ($\varphi_s + \varphi_i$) here is taken to be a constant, determined by best fit to the experimental QWSs. The results also agree reasonably with the ARPES results, as shown in Fig 4(d,h), although the slab calculation with the Si bilayer termination appears to agree best with experiment.

Previous studies on QWSs in Ag/Ge(111) [31][32], Ag/Si(111) [33] and Pb/Ge(111) [34] observed distortion of QWS dispersion and change of QWS bandwidth across the Si band edge, as a result of electronic coupling across the interface. Specifically, the QWSs outside the substrate band edge are fully confined within the film due to absent electronic coupling with substrates, while the QWS below the substrate band edge could couple to substrate bulk states, resulting in resonant QWS with larger bandwidth. At their anti-crossing point, the many-body interaction could give rise to kinks in the QWS band dispersion. Fig. 5(a,b) shows a large-scale ARPES spectra for 1.5 ML and 3 ML films, with Si band edge overlaid on top of the spectra. Indeed, QWSs outside the Si band edge appear to be better resolved compared to those below the band edge, although a sudden change in the QWS dispersion and bandwidth near the Si band edge could not be clearly identified, likely due to lack of crossing with sharp QWSs. Since the $CoSi_2$ lattice constant is well matched to the Si substrate, it would be natural to expect that the electronic coupling for QWSs below the Si band edge would be significant. Therefore, further studies are needed in the future to understand the electronic coupling across this coherently strained interface and the role of interfacial scattering.

3.3. Superconductivity of ultrathin $CoSi_2$ films

Bulk $CoSi_2$ is a Bardeen-Cooper-Schrieffer (BCS) superconductor with $T_c \approx 1.2$ K [9][35]. Therefore, it would be tempting to see if the electron-phonon coupling that leads to the superconductivity would affect the band dispersion near $E_F$ [36]. Fig. 5(c,d) shows the zoom-in view of the band dispersion near $E_F$ for 1 ML and 8 ML films. The high-lying optical phonons of $CoSi_2$ is at ~50 meV [37], but we do not find any clear signature of kink caused by electron-phonon coupling. Electron-phonon coupling may also appear as satellite peaks [38][39], if the electronic bandwidth is smaller than the phonon energy [36]. However, for $CoSi_2$ the bandwidth is much larger than the phonon energy. Therefore, satellite peaks are less likely to appear in this system [36] and indeed we do not observe such satellite peaks. Taken these into consideration, the electron-phonon coupling in this system might be very weak (given the small $T_c$), at least for the bands observed experimentally.

In previous transport measurements of the $CoSi_2$ films, the abrupt disappearance of superconductivity below about 10 nm was reported and a phenomenological explanation is the presence of a perturbed (or inter-mixed) layer of about 5 nm at the $Si/CoSi_2$ interface in which the electronic transport properties are dramatically altered compared to bulk [12]. We measured the transport properties of 100 ML and 10 ML $CoSi_2$ films (1 ML = 3.088 Å) and found that the superconductivity persists down to at least ~3 nm film (see Fig 6). The transition temperature is 1.09 K for 100 ML film and reduces to 0.73 K for 10 ML film. Here $T_c$ is defined to be the temperature where resistance drops to 10% of the normal-state value. The suppression of $T_c$ with decreasing film thickness has been observed in other QWS systems, such as Pb/Si(111) [6]. A magnetic field applied normal to the sample surface can suppress the superconductivity. The Werthamer-Helfand-Hohenberg (WHH) model [40] was applied to analyze the upper critical field. The obtained zero temperature value is ~1071 Oe for 100 ML and increases to ~1689 Oe for 10 ML, more than one order of magnitude higher than bulk. The increased upper critical field with reduced film thickness could be simply related to the reduced coherence length of the Cooper pairs. In addition, for ultrathin films, the inversion symmetry breaking at the interface could possibly induce spin-triplet paring component [41]. However, a clear identification of the spin-triplet component is difficult here, as the upper critical field is still small and orbital-limited. Nevertheless, probing superconductivity for even thinner films can be interesting (particularly for STM studies), as the interplay between quantum confinement and symmetry breaking at the interface may modify the superconductivity dramatically.

## 4. Conclusions

By combining the MBE growth, *in-situ* ARPES measurements and DFT calculations, we were able to track the evolution of the electronic structure of ultrathin epitaxial $CoSi_2$ films grown on Si (111) with increasing coverages. We directly observe well-defined QWSs, whose dispersion agree well with the DFT calculations. Our detailed analyses and calculations suggest that initial growth is dictated by the minimized surface energy, developing islands with preferred height of 2, 5 and 6 MLs. Afterwards, the surface energy oscillation becomes damped and the growth becomes atomically rough (but still epitaxial). The momentum-resolved electronic structure is overall consistent with the DFT calculations, indicating that the correlation effect from Co $3d$ electrons is likely weak. Our detailed comparison between ARPES and DFT also supported previous experiments that the epitaxial $CoSi_2$ films are terminated by additional Si bilayers on top. Finally, we found that superconductivity persists down to 3 nm, with a reduced $T_c$ but largely enhanced upper critical field. It would be interesting to understand how the superconductivity could be affected by the electronic quantization and the interfacial effect. For example, it would be particularly interesting to study the possible superconductivity of the 2, 5 and 6 ML magic thicknesses by STM or transport in the future.

After resubmitting this paper, we become aware of two recent arxiv papers [42, 43], which report evidence of triplet superconductivity in $CoSi_2$-based heterostructures.

**Acknowledgments**

We thank Zhiyong Nie, Prof. Michael Smidman and Prof. Yi Yin for helpful discussions. This work is supported by the National Science Foundation of China (No. 11674280), the National Key R&D Program of the MOST of China (Grant No. 2016YFA0300203, 2017YFA0303100).

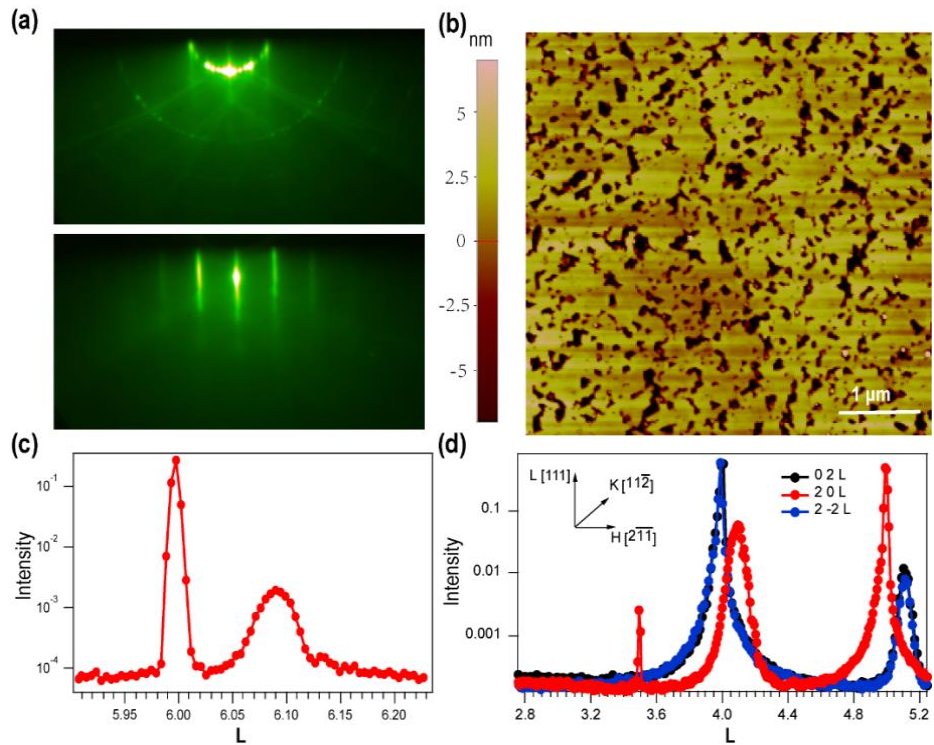

Fig 1. (a) RHEED patterns for the 7×7 reconstructed Si(111) surface (up) and after $CoSi_2$ film formation (down). (b) AFM image for a 20ML $CoSi_2$ film. (c) Scan along (H,K,L) = (0,0,L) for a 100ML $CoSi_2$ film from synchrotron X-ray scattering. (d) (0,2,L), (2,0,L) and (2,-2,L) scans for the same 100ML $CoSi_2$ film. The inset in (d) shows the definition of H,K,L directions in terms of bulk Si, which is normally used in surface X-ray scattering.

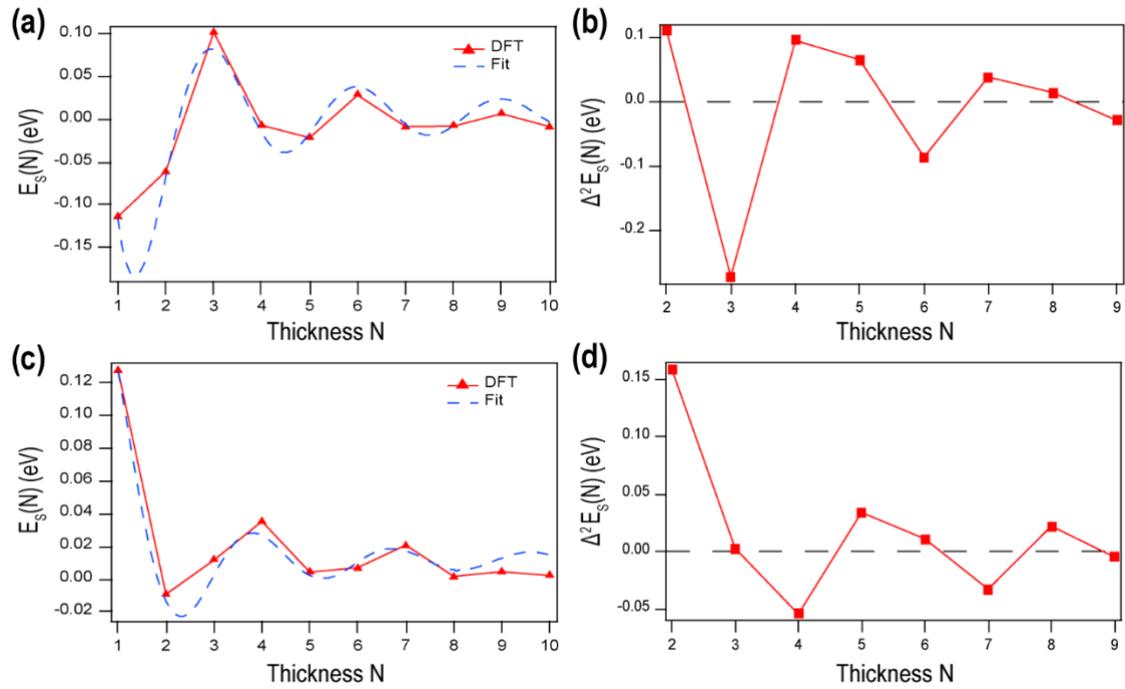

Fig 2. The calculated surface energy from DFT (a) and the second derivative of surface energy (b) for freestanding $CoSi_2$ films with bulk-like termination. (c,d) Same as (a,b), but for $CoSi_2$ films terminated by additional Si bilayers on both sides. The blue dashed lines in (a,c) are model fittings using eq. (1).

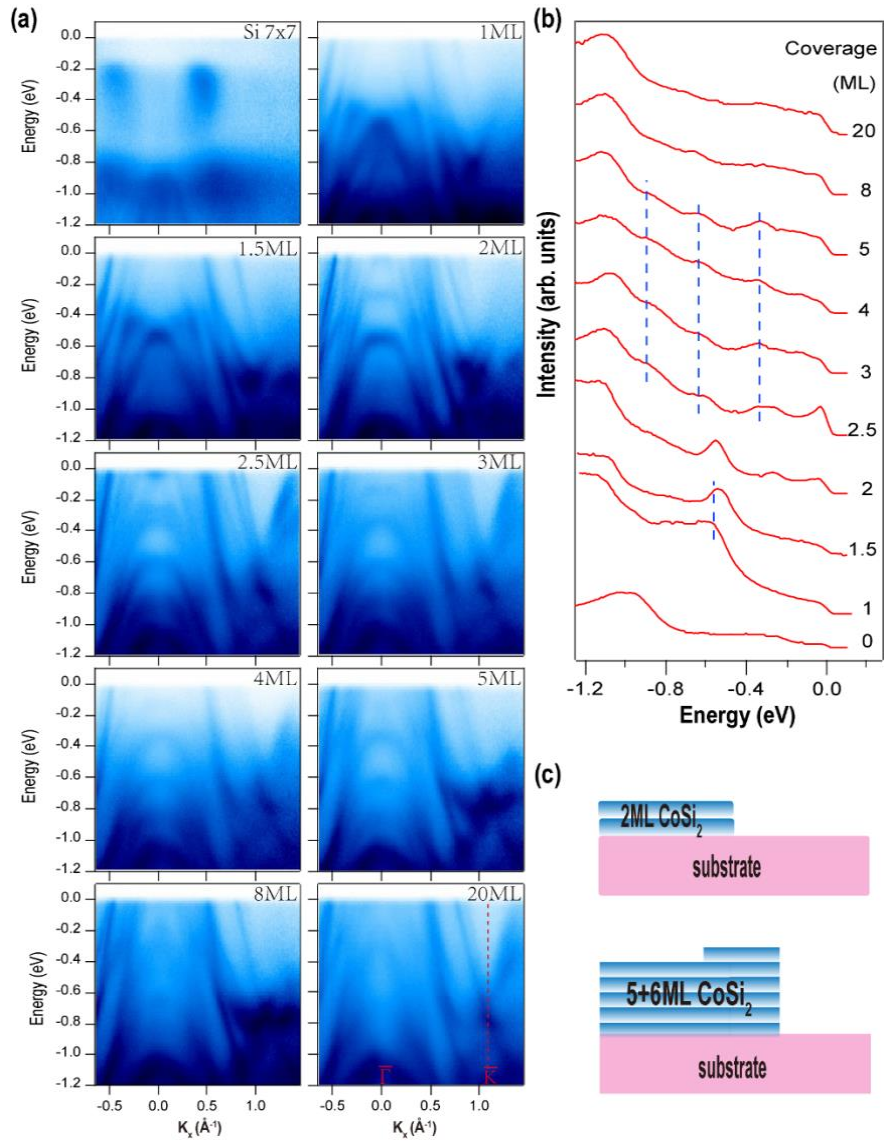

Fig 3. (a) Photoemission spectra from CoSi$_2$ films with different nominal coverages grown on Si(111). The photoemission data were taken along the $\bar{\Gamma}\bar{K}$ direction. (b) Evolution of EDCs at $\bar{\Gamma}$. Blue dashed lines are guide for eyes. (c) Schematic illustration of preferred island growth, i.e., 2 ML island at low coverages, 5 ML and 6 ML at intermediate coverages.

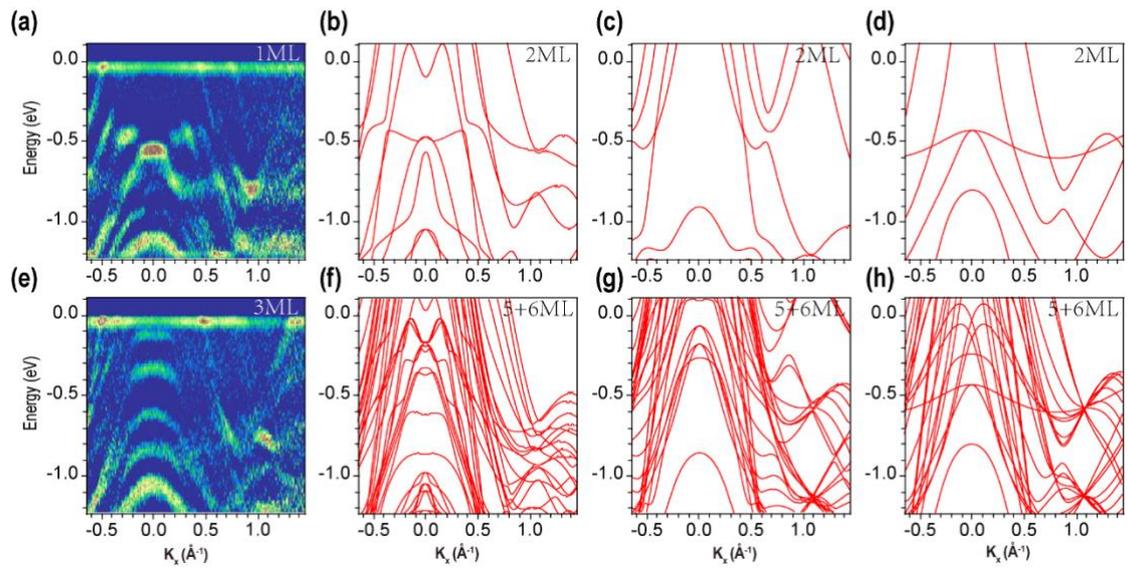

Fig 4. (a-d) Second derivative of ARPES data of the 1 ML film (nominal coverage) (a), in comparison with DFT calculations: slab calculation of 2 ML film with Si bilayer termination (b), slab calculation of 2 ML film with bulk-like termination (c), projected bulk bands (d). (e-h) same as (a-d), but for the 3 ML film (e). The DFT calculations (f-h) include contributions from both 5 ML and 6 MLs.

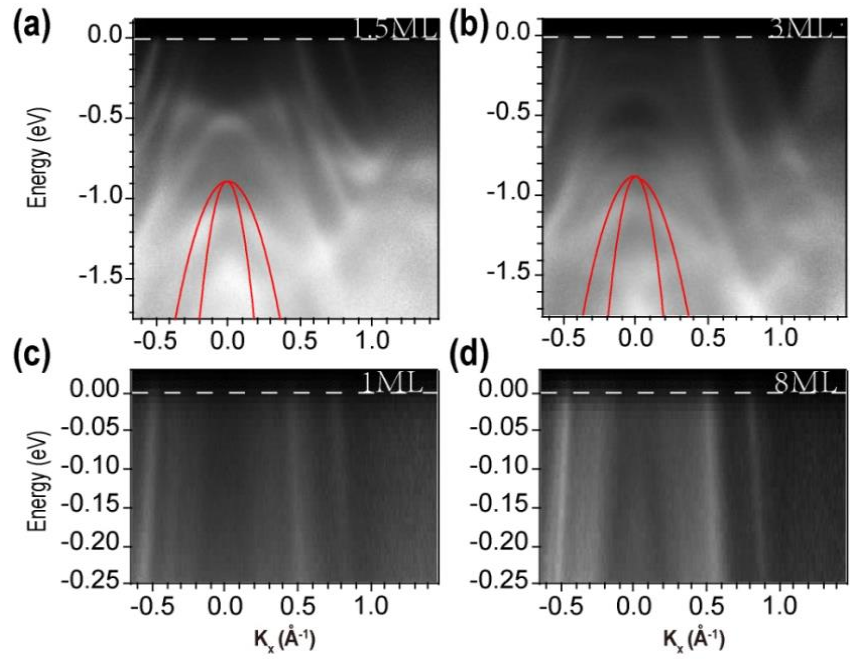

Fig 5. (a,b) Large-scale ARPES spectra of 1.5 ML and 3 ML films, with the Si band edge (red curves) overlaid on top. (c,d) Zoom-in view of the band dispersion near $E_F$ for 1 ML and 8 ML films.

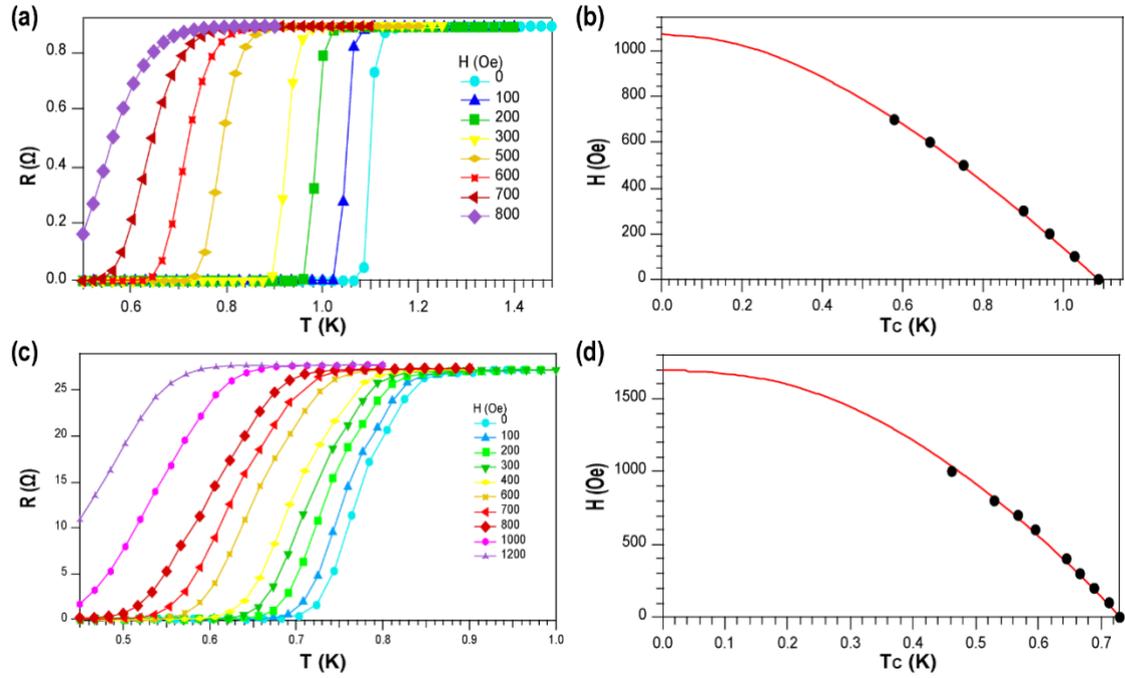

Fig 6. The temperature dependence of resistance for 100 ML (a) and 10 ML (c) $CoSi_2$ films under various magnetic fields. Upper critical field versus temperature for 100ML (b) and 10ML (d) films are extracted. The magnetic fields were applied along z direction. The fitted curves using WHH (red line) are also displayed.

# Supplementary information for
# Growth, Electronic Structure and Superconductivity of Ultrathin Epitaxial CoSi$_2$ Films

Yuan Fang[1,2,#], Ding Wang[1,2,#], Peng Li[1,2], Hang Su[1,2], Tian Le[1,2], Yi Wu[1,2], Guo-Wei Yang[1,2], Hua-Li Zhang[1,2], Zhi-Guang Xiao[1,2], Yan-Qiu Sun[2], Si-Yuan Hong[2], Yan-Wu Xie[2], Huan-Hua Wang[3], Chao Cao[4], Xin Lu[1,2,5], Hui-Qiu Yuan[1,2,5], Yang Liu[1,2,5,*]

[1]Center for Correlated Matter, Zhejiang University, Hangzhou, China

[2]Zhejiang Province Key Laboratory of Quantum Technology and Device, Department of Physics, Zhejiang University, Hangzhou, P. R. China

[3]Institute of High Energy Physics, Chinese Academy of Sciences, Beijing 100049, China

[4]Department of Physics, Hangzhou Normal University, Hangzhou, China

[5]Collaborative Innovation Center of Advanced Microstructures, Nanjing University, Nanjing, China

# these authors contribute equally to this work

* Email: yangliuphys@zju.edu.cn

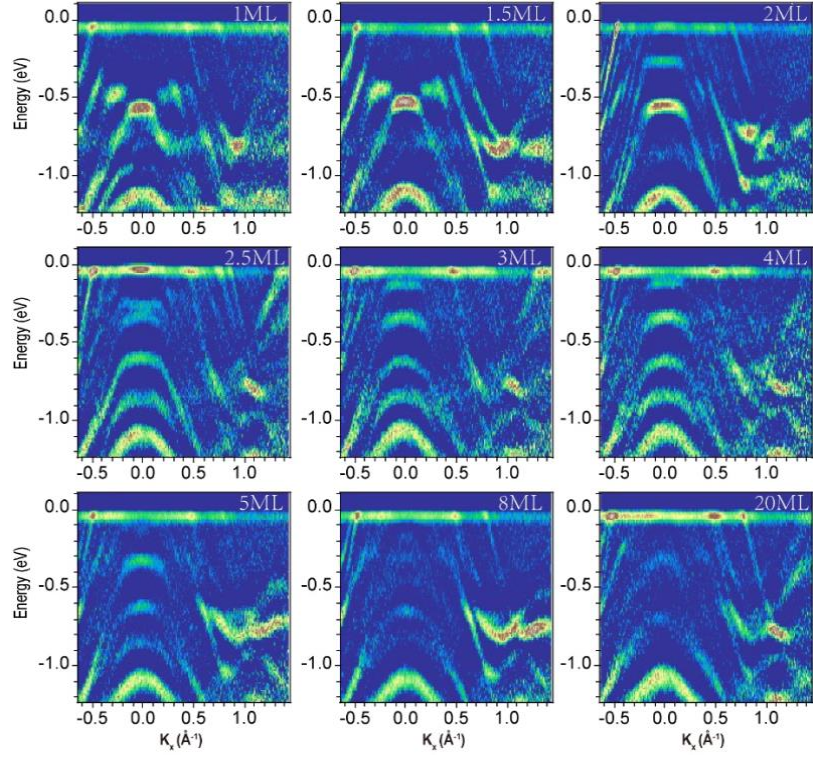

Figure S1. Second derivatives of ARPES spectra of $CoSi_2$ films with different thicknesses.

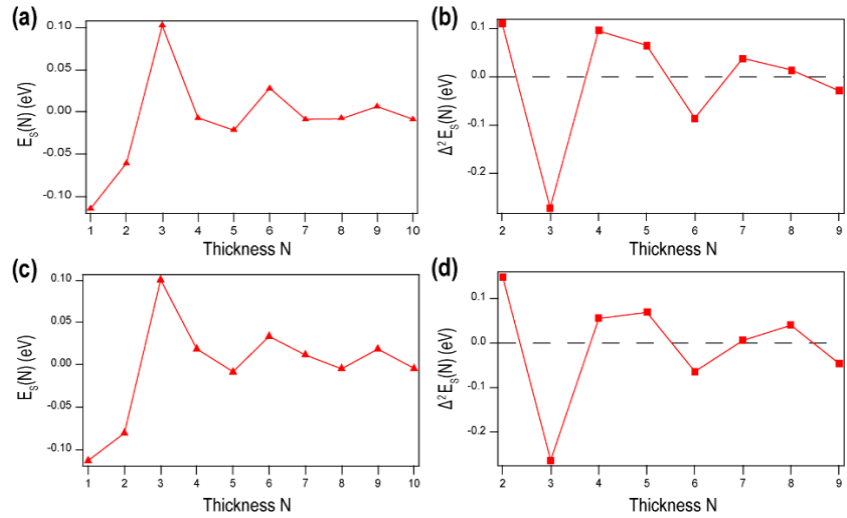

Figure S2. The calculated surface energy from DFT (a) and the second derivative of surface energy (b) for freestanding $CoSi_2$ films with bulk-like termination. The in-plane lattice constants are taken to be the same as Si substrate. (c,d) Same as (a,b), but the in-plane lattice constants are taken to be the same as bulk $CoSi_2$.